\documentclass[a4paper,11pt]{article}
\pdfoutput=1 

\usepackage{jcappub} 

\usepackage[T1]{fontenc} 
\usepackage{comment}
\usepackage{xcolor}

\title{
Closing the Cosmographic Hierarchy: Dynamical Attractors from Inflation to Reheating}

\author[a]{Seturumane Tema,}
\author[a, c]{S. Shajidul Haque,}
\author[b]{Saurya Das,}
\author[a, d, e]{and Peter Dunsby}

\affiliation[a]{Department of Mathematics and Applied Mathematics, University of Cape Town,\\ Cape Town-7700, South Africa}
\affiliation[b]{Department of Physics and Astronomy \& Quantum Horizons Alberta, University of Lethbridge, 4401 University Drive, Lethbridge AB T1K 3M4, Canada}
\affiliation[c]{National Institute for Theoretical and Computational Sciences (NITheCS),\\ Private Bag X1, Matieland,
South Africa}
\affiliation[d]{South African Astronomical Observatory, Observatory 7925, Cape Town, South Africa}
\affiliation[e]{Centre for Space Research, North-West University, Potchefstroom 2520, South Africa}
\emailAdd{tmxset001@myuct.ac.za}
\emailAdd{shajid.haque@uct.ac.za}
\emailAdd{saurya.das@uleth.ca}
\emailAdd{peter.dunsby@uct.ac.za}

\abstract{We develop a potential-independent cosmographic framework, in which cosmographic parameters are promoted to dynamical variables within a closed autonomous system. Although the cosmographic hierarchy is formally infinite, we achieve closure by mapping potential slow-roll parameters onto the kinematic phase space within General Relativity with a minimally coupled scalar field. Within this framework, we perform a stability analysis and show that inflationary (quasi–de Sitter) solutions arise as natural attractors, while stiff-fluid configurations act as repellers without invoking the slow‑roll approximation. To describe the transition to standard Big Bang evolution, we extend the system to include a radiation component and a phenomenological decay term. This leads to a generalized, potential-independent description of reheating characterized by an effective equation of state $w_{\text{eff}}$. We demonstrate that the radiation-dominated phase is the late-time attractor of the extended system. These results provide a unified kinematical description of the expansion history from inflation through reheating, bridging cosmography and scalar field dynamics.} 
\begin{document}
\maketitle

\flushbottom
\section{Introduction}
\label{sec:intro}

The theory of cosmic inflation suggests that, in the very early universe, within a tiny fraction of a second after the Big Bang, spacetime underwent a phase of rapid, quasi-exponential expansion \cite{Starobinsky:1980te,PhysRevLett.48.1437,Senatore:2010wk}. This brief yet profound epoch provides a compelling resolution to several long-standing problems in standard cosmology \cite{Guth:1980zm, Linde:1981mu, Albrecht:1982wi}. One of the most notable of these is the horizon problem: observations indicate that regions of the universe separated by vast distances exhibit nearly identical temperatures, despite being causally disconnected in the standard Big Bang framework \cite{Dicke:1965zz,Peebles:1994xt,Misner:1969hg}. Inflation resolves this apparent paradox by positing that these regions were once in causal contact prior to the rapid expansion, allowing thermal equilibrium to be established before being stretched to cosmological scales.\\

\noindent
A second issue concerns the observed near-flatness of the universe \cite{Planck:2018vyg,Eisenstein:2005su,Ryden:1970vsj,Carroll:2004st}. Within the standard cosmological model, even arbitrarily small deviations from spatial flatness in the early universe would grow with time, making the present-day near-flat geometry highly fine-tuned \cite{Baumann:2018muz,Kolb:1990vq}. The accelerated expansion during inflation dynamically drives the universe toward spatial flatness, thereby resolving this fine-tuning problem. In addition, various particle physics theories \cite{tHooft:1974kcl, Polyakov:1974ek, Preskill:1979zi} predict the production of heavy relics, such as magnetic monopoles, during early-universe phase transitions. The absence of such relics in observations is naturally explained by inflation, which exponentially dilutes their number density to negligible levels.\\ 

\noindent
In conventional realizations, inflation is typically driven by a scalar field, the inflaton, whose potential energy dominates the energy density of the universe. Under suitable slow-roll conditions \cite{PhysRevD.28.679,Martin:2013tda,Roest:2013fha,Peter:2013avv}, the scalar field evolves gradually, sustaining a prolonged phase of accelerated expansion. At the end of inflation, the inflaton decays, reheating the universe and initiating the standard hot Big Bang evolution. While this framework is phenomenologically successful, it relies on specific choices of scalar field potentials and initial conditions. This raises a key question: is inflation a consequence of specific scalar field models, or does it emerge more generally from the structure of cosmological dynamics? In this work, we address this question by adopting a potential-agnostic, cosmographic approach, where the expansion history is characterized directly through kinematical quantities, without committing to a specific form of the scalar field potential.\\

\noindent
We uncover several key results. First, within the closed cosmographic dynamical system, the de Sitter fixed point $(q, H) = (-1, H_*)$ is shown to be a robust stable attractor, while the stiff-fluid configuration ($q = 2$) acts as an unstable repeller. This implies that inflationary expansion emerges naturally from the dynamics, without fine-tuning of initial conditions or specific potential shapes. Second, when we relate the cosmographic parameters to inflationary observables, we find two algebraic branches for the jerk parameter. The ``minus branch'' gives rise to a red-tilted scalar spectral index ($n_s < 1$) compatible with Planck 2018 bounds on the tensor-to-scalar ratio $r$, whereas the ``plus branch'' is ruled out by its blue tilt. The quasi-de Sitter deformation parameter $\delta$ is constrained to the approximate range $10^{-3} \lesssim \delta \lesssim 10^{-2}$. Third, extending the system to include a radiation component and a decay term, we demonstrate that the radiation-dominated era ($q = 1$) is a global late-time attractor, providing a complete kinematical description from inflation through reheating to the hot Big Bang.\\

\noindent
The manuscript is organized as follows. In Section 2, we construct a closed two-dimensional dynamical system using the Hubble parameter $H$ and deceleration parameter $q$, and analyze its fixed points and stability. We then extend the system to three dimensions by promoting the jerk parameter $j$ to a dynamical variable. Section 3 connects the framework to observations by expressing inflationary observables in terms of cosmographic variables and identifying the physically viable branch. Sections 4 and 5 incorporate reheating through a two-fluid extension, first using a benchmark model and then generalizing to an effective equation of state. We conclude in Section 6 with a discussion of the implications and possible extensions of this framework.

\section{Framework and the Dynamical Closure Relations}
\label{sec:framework}
Cosmography provides a framework for describing cosmic expansion in terms of observable kinematic quantities, without specifying a particular form of the scalar field potential or modified gravity model 
\cite{Dunsby:2015ers, Weinberg:1972kfs, Busti:2015xqa, Sandage:1961zz}. By expressing the expansion history through the scale factor and its time derivatives, it offers a model-independent characterization of the background evolution. A natural question then arises: can such a kinematical framework be extended beyond reconstruction and series expansions to provide a self-consistent description of cosmic dynamics, with cosmographic parameters serving as fundamental phase-space variables?\\

\noindent
Although cosmographic quantities have been widely employed for phenomenological modelling and reconstruction \cite{rrnm-z3qy, chakraborty2024model, Goliath:1998na, Carloni:2004kp}, they are typically introduced only indirectly, rather than treated as dynamical variables of a closed autonomous system. The main obstruction is structural: cosmographic parameters form an infinite hierarchy. By definition, the Hubble, deceleration, jerk, snap and lerk parameters are
\begin{equation}
H = \frac{a^{(1)}}{a}, \quad
q = -\frac{a^{(2)}}{aH^2}, \quad
j = \frac{a^{(3)}}{a H^3}, \quad
s = \frac{a^{(4)}}{a H^4}, \quad
l = \frac{a^{(5)}}{aH^5}\;,
\end{equation}
where \(a(t)\) is the scale factor and \(a^{(n)}\) denotes the \(n^{th}\) derivative of \(a(t)\) with respect to the cosmic time. The evolution of each parameter introduces higher-order derivatives,
\begin{equation}
\dot q = H (2 q^2 + q - j), \quad
\dot j = H (\ldots \text{involving } s), \quad
\dot s = H (\ldots \text{involving } l), \dots
\end{equation}
\\
rendering the hierarchy infinite and preventing closure in terms of a finite set of variables without additional input. In this work, we overcome this obstruction by supplementing the cosmographic hierarchy with scalar field dynamics within General Relativity, thereby providing a physically motivated closure relation. This enables the construction of autonomous dynamical systems in which the cosmographic parameters themselves define the phase space.\\

\noindent
This framework offers two key advantages. First, it allows inflationary behaviour to be analysed directly at the level of kinematics within a closed cosmographic dynamical system. Second, the closure is achieved through scalar field dynamics in General Relativity without specifying a particular potential or invoking modified gravity. In this way, the approach bridges purely kinematical descriptions and full dynamical evolution, enabling phase-space analysis of features such as quasi--de Sitter trajectories and invariant sets directly in cosmographic space. Unlike most previous applications of cosmography, which focus on late-time expansion \cite{Bahamonde:2017ize,Vitagliano:2009sr}, our formulation applies cosmographic methods to the early universe, providing a kinematical perspective on inflationary dynamics.
\subsection{Scalar Field Reconstruction and Dynamical Closure}
\label{subsec:field_recon}
To provide a physical basis for closing the infinite cosmographic hierarchy, we consider a minimally coupled scalar field $\phi(t)$ evolving in a flat Friedmann-Lema\^itre-Robertson-Walker (FLRW) background. In units where ${M_\text{pl}}^{-2} = 8\pi G = 1$, the energy density $\rho_\phi$ and pressure $p_\phi$ of the inflaton are given by
\begin{equation}
\rho_\phi = \frac{1}{2} \dot{\phi}^2 + V(\phi), \qquad 
p_\phi = \frac{1}{2} \dot{\phi}^2 - V(\phi).
\end{equation}
The Friedmann and acceleration equations, $3H^2 = \rho_\phi$ and $\dot{H} = -\frac{(\rho_\phi + p_\phi)}{2}$, allow us to reconstruct the scalar field dynamics directly from the expansion history:
\begin{equation}\label{eq:field_recon}
\dot{\phi}^2 = -2 \dot{H}, \qquad 
V = 3H^2 + \dot{H}.
\end{equation}
Substituting the kinematic identity $\dot{H} = -H^2(1+q)$ into Eq.~\eqref{eq:field_recon}, the potential can be expressed in terms of cosmographic variables as $V = H^2(2 - q)$. To connect this reconstruction with inflationary dynamics, we introduce the potential slow-roll parameters, which characterize the flatness of the effective scalar field potential:
\begin{equation} \label {eq:epsilon_v}
\epsilon_V \equiv \frac{1}{2} \left( \frac{V_{,\phi}}{V} \right)^2 
\approx \frac{\dot{V}^2}{2 V^2 \dot{\phi}^2}, 
\qquad 
\eta_V \equiv \frac{V_{,\phi\phi}}{V} 
\approx \frac{\ddot{V} - V_{,\phi} \ddot{\phi}}{V \dot{\phi}^2}.
\end{equation}
Expressing the time derivatives of the potential in terms of $H$, $q$, $j$, and the snap parameter $s$, the slow-roll parameters can be written as the following algebraic relations:
\begin{align}\label{eq:slowroll}
\epsilon_V &= \frac{(-j + 3q + 4)^2}{4 (2 - q)^2 (1 + q)}, \\
\eta_V &= \frac{2j + 3q^2 + 18q + s + 12}{2 (2 - q)(1 + q)}.\label{slw}
\end{align}
Eq.~\eqref{eq:slowroll} therefore provides a critical closure relation for the two-dimensional cosmographic: system
\begin{equation}\label{eq:system}
\dot q = H (2 q^2 + q - j(q,\epsilon_V)), 
\qquad
\dot H = -H^2 (1 + q).
\end{equation}
\\
Solving this relation for $j$ truncates the hierarchy by expressing the jerk parameter in terms of the lower-order dynamical variable \(q\) and the slow-roll parameter \( \epsilon_V\).
\subsubsection*{Identification of fixed points}
\label{subsubsec:fixed_points}
The fixed points of the truncated system are determined by setting $\dot{q}=0$ and $\dot{H}=0$. From the condition $\dot{H}=0$, we find that the system settles either into a static state ($H=0$) or a state of constant expansion ($q=-1$). By solving the closure relation in Eq.~\eqref{eq:slowroll} for the jerk parameter $j$, we identify two distinct algebraic branches:
\begin{equation}\label{eq:j_eps}
j = 3q + 4 \mp 2(2 - q)\sqrt{1 + q}\sqrt{\epsilon_V}.
\end{equation}
Inserting Eq.~\eqref{eq:j_eps} into Eq.~\eqref{eq:system} reduces the two-dimensional cosmographic system to 
\begin{equation}\label{sfr}
\dot q = H \left[ 2 q^2 - 2q - 4 \pm 2 (2 - q) \sqrt{1 + q} \, \sqrt{\epsilon_V} \right],
\qquad
\dot H = -H^2 (1 + q).
\end{equation}
Substituting $q = -1$ into the expression of \(j\) in Eq.~\eqref{eq:j_eps} yields $j = 1$, which characterizes a pure de Sitter expansion phase consistent with $\Lambda$CDM \cite{Dunajski:2008tg}. The resulting fixed points of Eq.~\eqref{sfr} for both the minus and plus branches are summarized in the tables below. Note that while $F_1$ and $F_3$ represent static universes, $F_2$ remains the primary attractor of interest for early-universe dynamics.
\begin{table}[h!]
\centering
\begin{tabular}{ccc}
\hline
Fixed Point & $q$ & $H$ \\
\hline
$F_1$ & any real & 0 \\
$F_2$ & -1 & any real \\
\hline
\end{tabular}
\caption{Fixed points for the minus branch of the jerk parameter.}
\end{table}

\begin{table}[h!]
\centering
\begin{tabular}{ccc}
\hline
Fixed Point & $q$ & $H$ \\
\hline
$F_1$ & any real & 0 \\
$F_2$ & -1 & any real \\
$F_3$ & $\epsilon_V - 1$ & 0 \\
\hline
\end{tabular}
\caption{Fixed points for the plus branch of the jerk parameter.}
\end{table}
\subsection{Linear Stability Analysis (2D)}
\label{subsubsec:stability_2d}
To analyze the stability of the de Sitter fixed point $F_2$, we consider small linear perturbations $\delta$ such that $q = -1 + \delta$ with $\delta \ll 1$. In this neighborhood, the jerk parameter for the two branches expands as
\begin{equation}\label{kpert}
j \approx 1 + 3\delta \mp 6\sqrt{\delta\epsilon_V} + \mathcal{O}(\delta^{3/2}),
\end{equation}
with the corresponding derivative given by $\frac{\partial j}{\partial q} \approx \mp 3\frac{\sqrt{\epsilon_V}}{\sqrt{\delta}} + 3$. The Jacobian of the system, evaluated at the fixed point $H = H_*$, takes the form
\begin{equation}
J(-1+\delta, H_*) \simeq \begin{pmatrix} \pm 3H_* \frac{\sqrt{\epsilon_V}}{\sqrt{\delta}} - 6H_* & \pm 6\sqrt{\delta\epsilon_V} \\ -H_*^2 & -2H_*\delta \end{pmatrix}.
\end{equation}
The corresponding eigenvalues are
\[
\lambda_1 \approx \pm 3H_* \frac{\sqrt{\epsilon_V}}{\sqrt{\delta}} - 6H_*, 
\qquad 
\lambda_2 \approx -2H_*\delta,
\]
which govern the local dynamics. For the minus branch in an expanding universe ($H_* > 0$), the dominant eigenvalue $\lambda_1$ is negative, indicating that the de Sitter solution is a local attractor. In contrast, for the plus branch, $\lambda_1$ becomes positive and divergent as $\delta \to 0$, signaling that this branch is unstable and does not correspond to a physically viable inflationary trajectory. 
\subsection{Three-Dimensional Extension}\label{threeD}
While the two-dimensional system captures the essential dynamics near the de Sitter fixed point, a more complete description requires reinstating higher-order cosmographic variables. This naturally motivates extending the analysis beyond the reduced $(q,H)$ phase space by promoting the jerk parameter $j$ to a dynamical variable. To close the hierarchy, we express the snap parameter $s$ from Eq.~\eqref{slw} in terms of the potential slow-roll closure $\eta_V$:
\begin{equation}\label{kans}
s = 2 (2-q)(1+q)\, \eta_V - 2 j - 3 q^2 - 18 q - 12.
\end{equation}
This procedure truncates the infinite cosmographic series, resulting in a fully autonomous three-dimensional system:
\begin{equation}\label{eq:3D_system}
\begin{cases}
\dot H = - H^2 (1 + q),\\[1mm]
\dot q = H (2 q^2 + q - j),\\[1mm]
\dot j = H \Big[ s(q,j) + j (2 + 3 q) \Big] 
= H \Big[ 2 (2-q)(1+q) \, \eta_V - 2 j - 3 q^2 - 18 q - 12 + j (2 + 3 q) \Big].
\end{cases}
\end{equation}
In this extended framework, the fixed points of the 2D system can be embedded into the 3D phase space, with the additional dynamical degree of freedom allowing for richer trajectories and a more detailed characterization of the inflationary attractor. In particular, the de Sitter fixed point $F_2 =(q,j,H) = (-1,1,H_*)$ persists, while the system now also captures the evolution of the jerk parameter and its influence on the cosmic expansion history.
\subsubsection*{Linear Stability Analysis (3D)}
\begin{figure}[h!]
\centering
\includegraphics[width=0.7\textwidth]{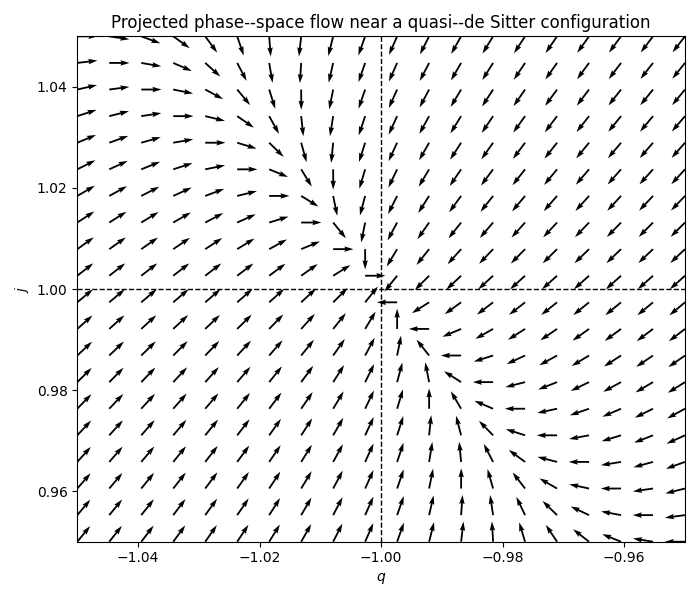}
\caption{Projection of the three-dimensional phase-space flow onto the $(q,j)$ plane at fixed $H=H_*$. Dashed lines indicate the invariant configurations $q=-1$ and $j=1$.}
\label{fig:phase_qj}
\end{figure}

We now extend the stability analysis to the three-dimensional autonomous system defined in Eq.~\eqref{eq:3D_system}. To this end, we consider small perturbations around the de Sitter fixed point \(F_2\),
\[
q = -1 + \delta q, \qquad
j = 1 + \delta j, \qquad
|\delta q|, |\delta j| \ll 1,
\]
with the Hubble parameter held fixed at \(H = H_* > 0\). Linearising the system in this neighbourhood results in the Jacobian
\begin{equation}
J(-1+\delta q,1+\delta j,H_*) =
\begin{pmatrix}
-2 H_* \delta q & - H_*^2 & 0 \\[1mm]
-3 \delta q - \delta j & -3 H_* + 4 H_* \delta q & -H_* \\[1mm]
0 & (6 \eta_V - 9) H_* & -3 H_* + 3 H_* \delta q
\end{pmatrix}.
\end{equation}
The eigenvalues of this linearised system, to leading order in \(\delta q\) and \(\delta j\), are
\[
\lambda_1 \simeq -3 H_*, \qquad
\lambda_2 \simeq -3 H_*, \qquad
\lambda_3 \simeq -2 H_* \delta q.
\]
For an expanding universe (\(H_* > 0\)) and small positive \(\delta q\), all eigenvalues are negative, confirming that the fixed point \(F_2= (-1,1,H_*)\) is locally stable in the full 3D phase space. The three-dimensional system eliminates the stability ambiguity present in the plus branch of the two-dimensional analysis by promoting \(j\) to a dynamical variable. The corresponding phase-space flow is illustrated in Fig.~\ref{fig:phase_qj}.  

\section{Observational Constraints and the Spectral Index}
\label{sec:obs_limits}
In this section, we confront the cosmographic dynamical system with observational data by expressing the standard inflationary observables such as the scalar spectral index $n_s$ and the tensor-to-scalar ratio $r$ in terms of the kinematic phase-space variables. This allows us to determine which region of the cosmographic parameter space is physically viable.\\

\begin{figure}[ht!]
    \centering
    \includegraphics[width=0.75\textwidth]{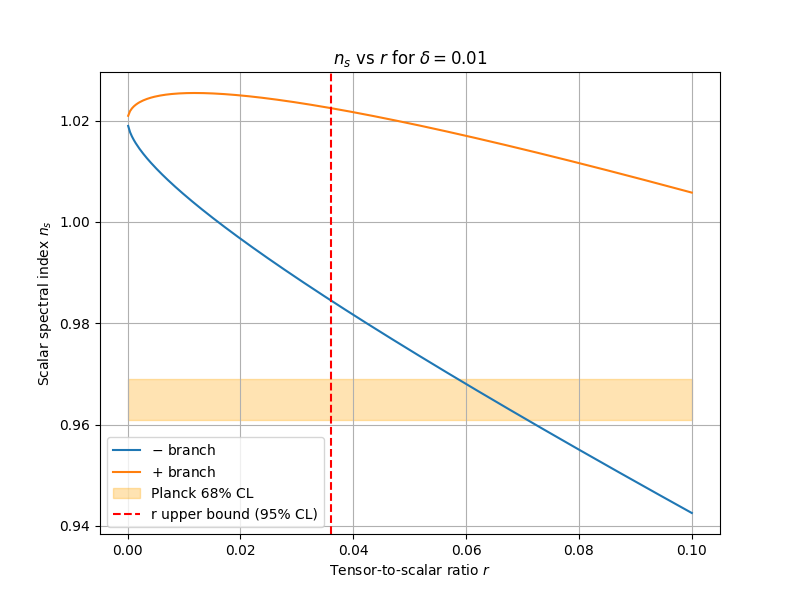} 
\caption{
Scalar spectral index $n_s$ as a function of the tensor-to-scalar ratio $r$ 
for quasi--de Sitter expansion with $\delta = 0.01$. 
The two curves correspond to the $\mp$ branches of Eq.~\eqref{eq:ns_r_final}.
The shaded orange region denotes the Planck 2018 $68\%$ confidence interval, 
$n_s = 0.965 \pm 0.004$, while the red dashed line indicates the 
approximate $95\%$ CL upper bound on $r$ ($r \lesssim 0.036$). 
Only the minus branch produces a red-tilted spectrum within the observationally allowed range of $r$; 
however, it does not intersect the Planck $68\%$ confidence region in this range.
}
\label{fig:nsr}
\end{figure}

\noindent
To leading order in the slow-roll approximation, the scalar spectral index is given by the standard relation:
\begin{equation}
n_{s}=1-6\epsilon_{V}+2\eta_{V}, \label{eq:ns_def}
\end{equation}
where $\epsilon_V$ and $\eta_V$ are the potential slow-roll parameters. To bridge this dynamical description with our kinematic framework, we utilize the relation for the jerk parameter $j$ that holds at leading slow-roll order:
\begin{equation}
j=1-3\eta_{V}+2\epsilon_{V}. \label{eq:j_sr}
\end{equation}
We work in the quasi-de Sitter regime near the fixed point $F_2$, where the parameters satisfy $\epsilon_V, \eta_V, \delta \ll 1$. In this neighborhood, the kinematic form of the jerk parameter was expanded as $j \approx 1+3\delta \mp 6\sqrt{\delta\epsilon_{V}}$. By equating the kinematic and dynamical expressions for $j$, we derive a consistency relation for $\eta_V$:
\begin{equation}
\eta_{V}=\delta \mp 2\sqrt{\delta \epsilon_{V}}-\frac{2}{3}\epsilon_{V}. \label{eq:eta_consist}
\end{equation}
\noindent
Substituting Eq.~\eqref{eq:eta_consist} into the definition of the spectral index in Eq.~\eqref{eq:ns_def}, we obtain $n_s$ as a function of the kinematic deviation $\delta$ and the slow-roll parameter $\epsilon_V$:
\begin{equation}
n_{s} \simeq 1+2\delta \mp 4\sqrt{\delta \epsilon_{V}}-\frac{22}{3}\epsilon_{V}.
\end{equation}
Using the leading-order relation for the tensor-to-scalar ratio, $r = 16\epsilon_V$, the spectral index can be expressed directly in terms of observable quantities:
\begin{equation}
n_{s} \simeq 1+2\delta \mp \sqrt{\delta r}-\frac{11}{24}r. \label{eq:ns_r_final}
\end{equation}
This relation reveals that deviations from scale invariance ($n_s \neq 1$) are governed by both the tensor-to-scalar ratio $r$ and the quasi-de Sitter deformation parameter $\delta$. As illustrated in Fig.~\ref{fig:nsr}, the square-root term in Eq.~\eqref{eq:ns_r_final} gives rise to two distinct algebraic branches. We compare these predictions with the Planck 2018 constraints ($n_s = 0.965 \pm 0.004$ and $r \le 0.036$).\\

\begin{figure}[ht!]
    \centering
    \includegraphics[width=0.75\textwidth]{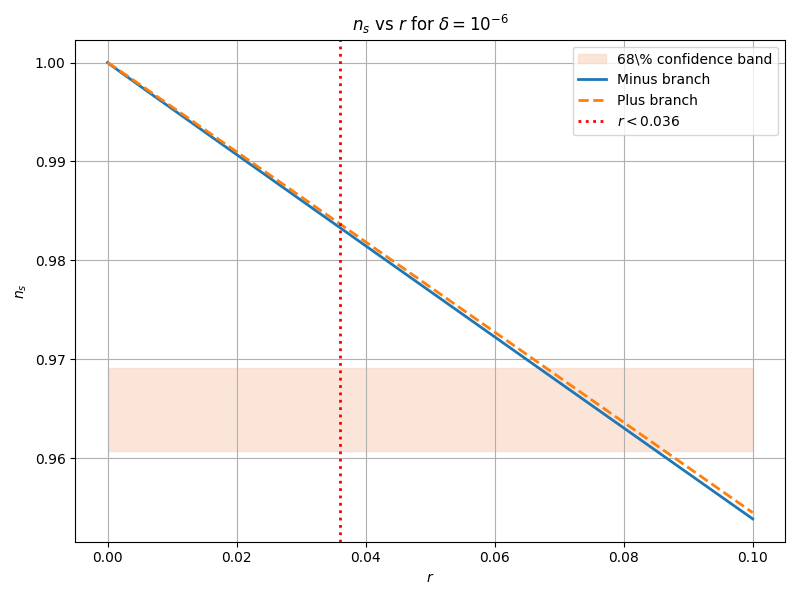} 
 \caption{Behaviour of the scalar spectral index $n_s$ for a deformation parameter $\delta = 10^{-6}$. 
At this value of $\delta$, the two solution branches merge and become indistinguishable. Both branches produce a red-tilted spectrum for values of $r$ within the observationally allowed interval; however, the resulting $n_s$ does not lie entirely within the Planck 2018 confidence region.
}
\label{fig:nsr_comparison}
\end{figure}

\noindent
Our analysis shows that when $\delta=0.01$ the plus branch leads to a blue-tilted spectrum ($n_s > 1$) for small $r$, which is strongly disfavoured by observations. In contrast, the minus branch results in a red-tilted spectrum ($n_s < 1$) for values of $r$ that lie within the observationally allowed range. However, the corresponding values of the spectral index $n_s$ lie outside the confidence region reported by Planck 2018, resulting in a persistent tension in the $(n_s, r)$ plane. To assess the robustness of this behaviour, we consider the limiting case $\delta = 10^{-6}$, for which the two solution branches coincide, as illustrated in Fig.~\ref{fig:nsr_comparison} and discussed in Section~\ref{threeD}. In this regime, both branches produce a red-tilted spectrum within the allowed range of $r$, yet the predicted values of $n_s$ remain outside the Planck confidence region. Taken together, these results indicate that, while the model successfully reproduces a red tilt in an observationally viable range of $r$, it does not achieve full consistency with current constraints due to a residual discrepancy in $n_s$. This type of mismatch is not uncommon in inflationary model building, where agreement with one observable does not guarantee simultaneous consistency with all datasets \cite{Freese:1990rb,Boubekeur:2005zm,Sse,Brandenberger:2011eq}, and suggests that further refinement of the deformation parameter $\delta$ may be required to achieve complete observational viability.\\

\noindent
To quantify this, we impose $n_s \simeq 0.965 \pm 0.004$ together with $r \le 0.036$. From Eq.~\eqref{eq:ns_r_final}, we have
\[
n_s - 1 + \frac{11}{24}r \simeq 2\delta \mp \sqrt{\delta r}.
\]
For representative values of $r$ in the allowed interval, $\frac{11}{24}r \lesssim 1.65\times 10^{-2}$, so that
\[
-3.5 \times 10^{-2} + \frac{11}{24}r \simeq 2\delta \mp \sqrt{\delta r}.
\]
Taking $r \sim 10^{-2}$ as a typical value gives
\[
-2 \times 10^{-2} \sim 2\delta \mp \sqrt{10^{-2}\delta}.
\]
Since the square-root term dominates for small $\delta$, consistency requires
\[
\sqrt{\delta r} \sim 10^{-2} \quad \Rightarrow \quad \delta r \sim 10^{-4}.
\]
Using $r \le 0.036$, this leads to a lower bound
\[
\delta \gtrsim \frac{10^{-4}}{0.036} \approx 3 \times 10^{-3}.
\]
Requiring that the linear contribution $2\delta$ does not significantly overshoot the Planck interval further suggests $\delta \lesssim 10^{-2}$. Hence, the model exhibits consistency at the level of order of magnitude estimates for
\[
10^{-3} \lesssim \delta \lesssim 10^{-2}.
\]
This provides an approximate constraint on the quasi--de Sitter deformation parameter $\delta$, identifying the range in which the model can achieve partial consistency with observational data.
\subsection{Global Dynamics and Inflationary Attractors}
In both the two and three-dimensional analyses, the Hubble parameter \(H\) was held fixed. As a consequence, the discussion was inherently local, effectively describing a single cosmological epoch namely, the quasi-de Sitter phase. To move beyond this limitation and capture the global evolution of the system, we now relax this assumption and allow \(\dot{H} \neq 0\). With this generalization, the dynamics reduce to a closed one-dimensional system of the form
\begin{equation}
\dot q = H\,F(q), \qquad F(q) = 2q^2 + q - j(q,\epsilon_V),
\end{equation}
where \(H>0\) corresponds to an expanding universe, and the jerk parameter \(j(q,\epsilon_V)\) is defined as in Eq.~\eqref{eq:j_eps}. The fixed points of the system are obtained by setting \(\dot q = 0\), which is equivalent to solving
\begin{equation}
F(q) = 2q^2 + q - j(q,\epsilon_V) = 0.
\end{equation}
Substituting the explicit form of the jerk parameter into this condition leads to
\begin{equation}
(q-2)\Big[(q+1)\mp \sqrt{1+q}\sqrt{\epsilon_V}\Big]=0,
\end{equation}
from which the fixed points can be identified as
\begin{equation}
q_* \in \{\,2,\; -1,\; \epsilon_V-1\ (\text{minus branch only})\}.
\end{equation}

To assess the nature of these fixed points, we consider the linear stability of the system. Since \(H>0\), it merely rescales the time variable and does not influence stability. The behaviour near a fixed point \(q_*\) is therefore governed by the sign of \(F'(q_*)\), with stability corresponding to \(F'(q_*)<0\) and instability to \(F'(q_*)>0\). Differentiating \(F(q)\) with respect to \(q\), we obtain
\begin{equation}
F'(q)=4q+1 - j'(q),
\end{equation}
where
\begin{equation}
j'(q)=3 \mp 2\sqrt{\epsilon_V}
\left[
-\sqrt{1+q}+\frac{2-q}{2\sqrt{1+q}}
\right].
\end{equation}
We note that the presence of the square root restricts the phase space to \(q \ge -1\). We begin with the inflationary fixed point at \(q=-1\). Writing \(q=-1+\delta\) with \(\delta \ll 1\), one finds that \(j'(q)\) diverges as \(\delta \to 0\), rendering the linear analysis formally singular at this point. Consequently, stability must instead be determined by examining the sign of \(F(q)\) in a neighbourhood of \(q=-1\). This analysis shows that the minus branch results in \(F<0\), indicating stability, whereas the plus branch gives \(F>0\), corresponding to instability. Next, we consider the stiff-fluid point at \(q=2\). In the regime \(\epsilon_V \ll 1\), one finds \(j'(2)\approx 3\), which implies
\begin{align}
    F'(2)=4(2)+1-3=6>0,
\end{align}
establishing that this fixed point is unstable. Finally, we examine the quasi-de Sitter solution \(q=\epsilon_V-1\), which exists only on the minus branch. For small \(\epsilon_V\), one finds \(F'(q_*)<0\), demonstrating that this point is locally stable. These results are summarised below:
\begin{center}
\begin{tabular}{c c c}
\hline
Fixed point & Stability & Interpretation \\
\hline
$q=-1$ & stable (minus branch) & inflation \\
$q=\epsilon_V-1$ & stable & quasi-de Sitter \\
$q=2$ & unstable & stiff fluid \\
\hline
\end{tabular}
\end{center}
The phase‑space analysis of the one‑dimensional system reveals that the stiff‑fluid fixed point at \(q=2\) is a repeller, while the de Sitter point at \(q=-1\) is a stable attractor. This structure naturally suggests that if the universe began in a stiff‑fluid dominated phase, it would inevitably evolve toward the inflationary attractor. Such a scenario is not merely a mathematical curiosity; it has been explored in concrete cosmological models \cite{Copeland:1997et, Peebles:1987ek,Liddle:1998xm}. Stiff fluids generally appear in pre‑Big‑Bang scenarios and in some bouncing cosmologies as a high‑curvature phase preceding the bounce \cite{Gasperini:1992em, Damour:2002et,Spokoiny:1993kt,PhysRevD.92.103004}. In our framework, the repelling nature of the stiff‑fluid fixed point guarantees that any trajectory near \(q=2\) will be driven toward the de Sitter attractor, provided the universe expands (\(H>0\)). This provides a dynamical explanation for why inflation is a robust outcome even if the early universe started in a high‑pressure, kinetically dominated state.
\subsection{Evolution of the Cosmographic Flow: From Inflation to Reheating}
The global phase-space structure reveals a clear dynamical flow: trajectories are repelled from the stiff-fluid regime and attracted toward quasi–de Sitter configurations. In this sense, accelerated expansion arises dynamically as an attractor of the closed cosmographic system, rather than being imposed through specific initial conditions or slow-roll assumptions. To describe the subsequent transition to the standard hot Big Bang phase, we extend the framework to include reheating. This requires introducing additional physics beyond the single scalar field description, since a lone field in a stable attractor cannot produce a radiation-dominated universe.
We therefore incorporate a second component that represents radiation, together with a phenomenological decay term $\Gamma$ that transfers energy from the inflaton to the radiation bath. This extension allows the system to evolve from the inflationary attractor through a post-inflationary oscillatory phase and ultimately toward radiation domination.
In the following sections, we develop this two-fluid framework in detail. We first consider a benchmark scenario based on a quadratic potential ($V \propto \phi^2$) to illustrate the reheating mechanism and then generalize the analysis using an effective equation of state $w_{\text{eff}}$ to maintain potential independence.

\section{Reheating: Extending to Two Fluids}
\label{reheating_1}
The above analysis describes a single scalar field. Without additional physics, the post-inflationary evolution cannot lead to radiation domination. For a quadratic potential, the oscillating inflaton behaves as pressureless matter (\(w=0\)) on average, giving \(q=\frac{1}{2}\). For a quartic potential, it behaves as radiation (\(w=\frac{1}{3}\)) giving \(q=1\), but such potentials are disfavored. In any case, a single field cannot produce a radiation-dominated universe without decaying. To model reheating, we introduce a radiation component with energy density \(\rho_r\) and pressure \(p_r=\frac{\rho_r}{3}\), and include a decay term \(\Gamma\) representing the inflaton's decay into radiation. The equations are
\begin{equation}
\dot{\rho}_\phi + 3H(\rho_\phi + p_\phi) = -\Gamma \rho_\phi,\qquad 
\dot{\rho}_r + 4H\rho_r = \Gamma \rho_\phi,
\end{equation}
with \(H^2 = \frac{\rho_\phi+\rho_r}{3}\). We assume that after inflation the inflaton oscillates in a quadratic potential, so on average \(p_\phi=0\) (matter-like behavior). This is the simplest reheating model.

\subsection{Dynamics of Reheating and the Transition to Radiation}
\label{sec:reheating_dynamics}
To characterize the transition from inflation to the standard thermal history of the Big Bang, we extend the cosmographic framework to a two-fluid dynamical system. We define the dimensionless density parameters for the radiation component and the scalar field as
\begin{equation}
\Omega_r = \frac{\rho_r}{3H^2}, \qquad \Omega_\phi = \frac{\rho_\phi}{3H^2},
\end{equation}
subject to the constraint $\Omega_\phi + \Omega_r = 1$. By adopting the number of e-folds $N = \ln a$ as the independent time variable, the evolution of the system can be tracked via the logarithmic derivative $\frac{d}{dN} = (\frac{1}{H})\frac{d}{dt}$. The dynamical evolution of the radiation density is then captured by the relation
\begin{equation}\label{eq:log_deriv_omega}
\frac{d\ln\Omega_r}{dN} = \frac{d\ln\rho_r}{dN} - 2\frac{d\ln H}{dN}.
\end{equation}
The individual terms in this expression are determined by the conservation equation for radiation, 
\begin{equation}
\dot{\rho}_r = -4H\rho_r + \Gamma\rho_\phi, 
\end{equation} 
and the definition of the deceleration parameter, $q = -1 - \frac{\dot{H}}{H^2}$. For this two-fluid system, the deceleration parameter is related to the energy content by $q = \frac{1}{2} (1+ \Omega_r)$, which implies that $1+q = \frac{(3+\Omega_r)}{2}$. Consequently, the Hubble evolution is governed by 
\begin{equation}\frac{d\ln H}{dN} = -\frac{(3+\Omega_r)}{2}.\end{equation}
Substituting these components into Eq.~\eqref{eq:log_deriv_omega} and multiplying through by $\Omega_r$, we arrive at the evolution equation for the radiation density parameter:
\begin{equation}\label{eq:omega_r_evolution}
\frac{d\Omega_r}{dN} = (1-\Omega_r)\left( \frac{\Gamma}{H} - \Omega_r \right).
\end{equation}
Together with the expansion rate evolution,
\begin{equation}\label{eq:H_evolution}
\frac{1}{H}\frac{dH}{dN} = -\frac{3+\Omega_r}{2},
\end{equation}
these expressions form a closed autonomous system. \\

\noindent
The fixed points of this system are identified by the condition $\frac{d\Omega_r}{dN} = 0$, which yields the radiation-dominated solution $\Omega_r = 1$. The stability of this state is assessed by considering a small perturbation $\epsilon$ such that $\Omega_r = 1 - \epsilon$. In the limit $\epsilon \ll 1$, the linear evolution is given by $\frac{d\epsilon}{dN} = -\epsilon(\frac{\Gamma}{H} - 1)$. As the expansion progresses and $H$ decreases, the ratio $\frac{\Gamma}{H}$ inevitably grows; once the reheating condition $\frac{\Gamma}{H} > 1$ is satisfied, the perturbation decays, establishing $\Omega_r = 1$ as a stable late-time attractor.\\

\noindent
The physical significance of this transition is underscored by the evolution of the cosmographic parameters $q$ and $j$. The complete cosmic trajectory begins at the inflationary attractor ($q \approx -1$), proceeds through the post-inflationary oscillatory phase where the inflaton behaves as pressureless matter ($q = \frac{1}{2}, j=1$), and finally settles at the radiation-dominated fixed point ($q = 1, j = 3$). This unified kinematical description effectively bridges the gap between the high-energy inflationary epoch and the subsequent radiation-dominated era of the standard Big Bang model.\\

\noindent
While the preceding analysis successfully captures the reheating transition, it relies on the specific assumption of a quadratic potential where the inflaton behaves as pressureless matter $(w=0)$ on average. To ensure our cosmographic framework remains truly model-independent, we must move beyond specific potential forms. In the following section, we generalize this description by introducing an effective equation of state, $w_{\rm eff}$, to characterize the oscillating inflaton. This approach allows us to describe the decay into radiation and the subsequent approach to the late-time attractor without tying the dynamics to a particular scalar field geometry.
\section{Generalized Reheating: Two Fluids with an Effective Equation of State}
\label{sec:generalized_reheating}
While the preliminary analysis of reheating often assumes a specific scalar field potential, a truly model-independent cosmographic framework must be capable of describing the transition to radiation regardless of the underlying inflaton geometry. The single-field description of the early universe is fundamentally insufficient for this task, as a lone scalar field in a stable attractor regime cannot, by itself, initiate the standard radiation-dominated era. To resolve this, we generalize our dynamical system to include a radiation component with energy density $\rho_r$ and pressure $p_r = \frac{\rho_r}{3}$.\\ 

\noindent
The transfer of energy from the inflaton to the thermal bath is modeled via a phenomenological decay term, $\Gamma$. In this generalized regime, the oscillating inflaton is no longer tied to a specific potential; instead, it is characterized by an effective equation of state parameter $w_{\text{eff}}$, which represents the average pressure-to-density ratio $\frac{\langle p_\phi \rangle} {\langle \rho_\phi \rangle}$ over many oscillations. This parameter effectively encodes the local shape of the potential minimum. For instance, a quadratic minimum corresponds to $w_{\text{eff}} = 0$ (behaving as pressureless matter), whereas a quartic minimum yields $w_{\text{eff}} = \frac{1}{3}$. More complex potentials, such as those found in axion monodromy models, can yield a wider spectrum of effective values. The coupled conservation equations for this two-fluid system are expressed as:
\begin{equation}
\dot{\rho}_\phi + 3H(1+w_{\text{eff}})\rho_\phi = -\Gamma \rho_\phi, \qquad \dot{\rho}_r + 4H\rho_r = \Gamma \rho_\phi,
\end{equation}
with the expansion rate governed by the Friedmann equation $H^2 = \frac{\rho_\phi + \rho_r}{3}$.
\subsection{Dimensionless Variables and Evolution Equations}
To facilitate a dynamical systems analysis, we transform the physical densities into dimensionless variables that represent the relative energy contribution of each component. We define the density parameters for radiation and the scalar field as:
\begin{equation}
\Omega_r = \frac{\rho_r}{3H^2}, \qquad \Omega_\phi = \frac{\rho_\phi}{3H^2},
\end{equation}
subject to the physical constraint $\Omega_\phi + \Omega_r = 1$. To track the evolution of the system over cosmological timescales, we employ the number of e-folds $N = \ln a$ as our independent time variable. This leads to the operational derivative $\frac{d}{dt} = H \frac{d}{dN}$, which simplifies the analysis of the cosmic flow through the phase space.\\

\noindent
The core of our generalized reheating model lies in the evolution of the radiation density parameter $\Omega_r$. By taking the logarithmic derivative of its definition, we obtain:
\begin{equation}\label{eq:gen_log_omega}
\frac{d\ln\Omega_r}{dN} = \frac{d\ln\rho_r}{dN} - 2\frac{d\ln H}{dN}.
\end{equation}
The first term in Eq.~\eqref{eq:gen_log_omega} is derived directly from the radiation conservation equation, yielding:
\begin{equation}
\frac{d\ln\rho_r}{dN} = \frac{\dot{\rho}_r}{H\rho_r} = -4 + \frac{\Gamma}{H}\frac{\Omega_\phi}{\Omega_r}. 
\end{equation}
The second term, representing the change in the Hubble expansion rate, is determined by the total pressure of the two-fluid mixture. Using the Friedmann equations, we find:
\begin{equation}\label{eq:gen_h_dot}
\frac{\dot{H}}{H^2} = -\frac{1}{2}\left[ 3(1+w_{\text{eff}})\Omega_\phi + 4\Omega_r \right].
\end{equation}
Substituting these relations into the logarithmic derivative and simplifying the non-decay contributions, we arrive at the following autonomous evolution equation for the radiation density:
\begin{equation}\label{eq:gen_omega_compact}
\frac{d\Omega_r}{dN} = (1-\Omega_r)\left( \frac{\Gamma}{H} - (1-3w_{\text{eff}})\Omega_r \right).
\end{equation}
Simultaneously, the evolution of the expansion rate $H$ is decoupled and governed by:
\begin{equation}
\frac{1}{H}\frac{dH}{dN} = -\frac{1}{2}\left[ 3(1+w_{\text{eff}}) + \Omega_r(1-3w_{\text{eff}}) \right].
\end{equation}
Together, these equations provide a comprehensive description of the system's trajectory from the end of the inflationary epoch to the completion of reheating.
\subsection{Fixed Points and Stability}
An analysis of Eq.~\eqref{eq:gen_omega_compact} reveals the critical points that define the late-time behavior of the universe. The condition $\frac{d\Omega_r}{dN} = 0$ is satisfied by the pure radiation solution $\Omega_r = 1$, or by the equilibrium configuration \(
\Omega_r = \Gamma/\bigl[H(1 - 3w_{\mathrm{eff}})\bigr]
\). Because $H$ is a strictly decreasing function of time in an expanding universe, the latter is not a static fixed point but rather a transient state. \\

\noindent
To determine the stability of the radiation-dominated state, we consider a linear perturbation $\epsilon$ such that $\Omega_r = 1 - \epsilon$. In the limit $\epsilon \ll 1$, the linearized evolution is expressed as:
\begin{equation}
\frac{d\epsilon}{dN} \approx -\epsilon\left( \frac{\Gamma}{H} - (1-3w_{\text{eff}}) \right).
\end{equation}
As $H$ continues to decrease, the ratio $\frac{\Gamma}{H}$ grows. Once $\frac{\Gamma}{H}$ exceeds the threshold $(1-3w_{\text{eff}})$, the perturbation $\epsilon$ begins to decay exponentially. This identifies $\Omega_r = 1$ as the global late-time attractor for any $w_{\text{eff}} < \frac{1}{3}$. This condition generalizes the standard reheating scenario, demonstrating that a radiation-dominated Big Bang is the inevitable outcome of the dynamical cosmographic flow.
\subsection{Connection to Cosmographic Parameters}
The final step in our generalized analysis is to relate the energy-density evolution back to the kinematic parameters $q$ and $j$. Using the relation derived from the Friedmann equations, the deceleration parameter for the two-fluid system is:
\begin{equation}
q = -1 + \frac{1}{2}\left[ 3(1+w_{\text{eff}}) + \Omega_r(1-3w_{\text{eff}}) \right].
\end{equation}
This allows us to map the entire expansion history onto a single cosmographic trajectory. During the inflationary epoch, the scalar field dominance ensures $q \approx -1$. Following the end of inflation, the universe enters the oscillatory phase where $q = \frac{1}{2}(1+3w_{\text{eff}})$ (assuming $\Omega_r \approx 0$). Finally, as reheating concludes and $\Omega_r \to 1$, the deceleration parameter reaches the stable value of $q=1$. The jerk parameter follows a similar transition, moving from its inflationary attractor value of 1 to the radiation-dominated value of 3. This unified framework demonstrates that the progression from inflation to the Big Bang is not an artifact of specific potentials, but a robust consequence of the dynamical attractors inherent in cosmographic space.
\section{Conclusion}
\label{sec:conclusion}
In this work, we have developed a self-consistent framework for describing the early universe by promoting cosmographic parameters specifically the Hubble parameter $H$, the deceleration parameter $q$, and the jerk parameter $j$ to dynamical variables within a closed autonomous system. While the cosmographic hierarchy is intrinsically infinite, we achieved closure by introducing relations derived from scalar field dynamics in General Relativity, without specifying a particular form of the scalar field potential.\\

\noindent
Our dynamical analysis identifies the de Sitter expansion phase, defined by the fixed point $q = -1$ and $j = 1$, as a robust stable attractor in both two-dimensional and three-dimensional phase-space extensions. Conversely, we have demonstrated that high-pressure ``stiff-fluid'' configurations ($q = 2$) act as unstable repellers. This results in a dynamical flow where trajectories are naturally repelled from the stiff-fluid state and attracted toward inflationary configurations, suggesting that accelerated expansion emerges as a generic feature of the expansion history rather than an artifact of fine-tuned initial conditions.\\

\noindent
Furthermore, we test the model against observational constraints by recasting the scalar spectral index $n_s$ and the tensor-to-scalar ratio $r$ in terms of cosmographic parameters. The resulting relation exhibits two algebraic branches associated with the jerk parameter. The ``plus branch'' generically leads to a blue-tilted spectrum ($n_s > 1$) and is therefore disfavoured by observations. In contrast, the ``minus branch'' produces a red-tilted spectrum ($n_s < 1$) for values of $r$ within the observationally allowed range, making it phenomenologically more viable. In the regime of a very small deformation parameter ($\delta \ll 1$), the two branches become effectively indistinguishable as the square-root term vanishes, leading to a degenerate prediction for $n_s$. This degeneracy does not resolve the tension with observations. Nevertheless, the agreement remains incomplete, as no choice of the deformation parameter $\delta$ allows for simultaneous consistency with the full Planck 2018 $(n_s, r)$ confidence region within the present approximation. Achieving a red-tilted spectrum nevertheless requires a small deviation from an exact de Sitter phase, corresponding to a quasi--de Sitter deformation parameter in the approximate range $10^{-3} \lesssim \delta \lesssim 10^{-2}$.\\

\noindent
To provide a complete description of the cosmic evolution, we extended the single-fluid system to a two-fluid dynamical framework incorporating a radiation component and a phenomenological decay term $\Gamma$. This generalization allows for a potential-independent treatment of reheating characterized by an effective equation of state $w_{eff}$. Our global stability analysis confirms that the radiation-dominated era ($q = 1$) serves as the global late-time attractor of the system once the inflaton decays.\\

\noindent
Ultimately, these results provide a unified kinematical description of the expansion history that bridges the gap between purely phenomenological cosmography and fundamental scalar field dynamics. By characterizing the critical epochs of the early universe as natural attractors in cosmographic space, this framework offers a self-consistent account of cosmic evolution from the onset of inflation through the matter-dominated oscillatory phase to the final radiation-dominated era. \\

\noindent
The present work opens several avenues for extending the cosmographic dynamical approach to cover the entire history of the universe, from a possible pre‑inflationary bounce to the late‑time accelerating phase. The current analysis is restricted to expanding cosmologies (\(H>0\)). However, many quantum gravity and string‑inspired scenarios predict a bouncing phase that replaces the initial singularity \cite{Brandenberger:2016vhg,Burger:2018hpz,Das:2024rop}. In such models, the universe contracts from a large, cold state, bounces at a minimum scale factor, and then expands through a stiff‑fluid phase (often with \(w=1\)) into inflation. Our cosmographic closure relations, derived from general relativity, are valid for any sign of \(H\) as long as the FLRW metric and the scalar‑field equations hold. By allowing \(H\) to become negative during contraction, one can extend the dynamical system to include bounces. The fixed point \(q=2\) (stiff fluid) would then play a dual role: as a repeller during expansion, it would drive the universe toward inflation; during contraction, it could act as an attractor, describing a kinetic‑dominated contraction phase \cite{Lehners:2008vx}. A natural future project is to construct a global phase space that includes both \(H>0\) and \(H<0\) branches, and to study whether the bounce can be realised as a smooth transition through a hypersurface \(H=0\), possibly with a modified closure condition near the bounce.\\

\noindent
A related direction concerns ekpyrotic and slow‑contraction scenarios, which involve a contracting universe with a scalar field having a stiff equation of state (\(w\gg 1\)), corresponding to \(q\gg 2\) – far from the inflationary attractor identified here. Our analysis, restricted to expansion (\(H>0\)), does not directly apply to such phases. However, a straightforward extension of our dynamical system to the contracting branch (\(H<0\)) could determine whether an ekpyrotic phase can be a robust attractor in its own right, or whether it is dynamically disfavoured. This would complement ongoing work on bouncing cosmologies and provide a more complete cosmographic picture of the very early universe.\\

\noindent
Our two‑fluid generalisation already demonstrates that the radiation dominated era (\(q=1\)) is a late‑time attractor after reheating. However, current observations indicate that the universe is now entering another accelerated phase, often attributed to a cosmological constant or dark energy, corresponding to \(q\approx -1\) again. In the language of our cosmographic phase space, this is a second de Sitter attractor. A natural extension is to include a third fluid component (e.g., a cosmological constant or a quintessence field) with an effective equation of state \(w_{\mathrm{DE}}\approx -1\). The corresponding dynamical system would then have two stable de Sitter fixed points: one at early times (inflation) and one at late times (dark energy). The challenge is to reconcile the fact that the universe must evolve from the early de Sitter attractor through a long matter‑dominated phase to the late de Sitter attractor. This would require a careful analysis of the intermediate fixed points and the possible existence of heteroclinic orbits connecting them. The same cosmographic closure technique could be applied to a scalar field with a potential that drives both early and late acceleration, such as the Starobinsky model or certain \(\alpha\)-attractors \cite{Kallosh:2013hoa, Copeland:2006wr,Aviles:2012ay}. The effective equation of state parameter \(w_{\mathrm{eff}}\) introduced in Section 5 can be promoted to a time‑dependent function that transitions from \(w_{\mathrm{eff}}\approx -1\) (inflation) to \(0\) (matter) to \(\frac{1}{3}\) (radiation) and finally back to \(-1\) (dark energy). Analysing such a flow in the extended phase space of cosmographic parameters would provide a unified dynamical description of the entire cosmic history. 
More broadly, the dynamical systems approach developed here offers a flexible framework for studying cosmological evolution directly in terms of observable kinematic quantities. By combining cosmography with physically motivated closure relations, it provides a bridge between phenomenological reconstruction and fundamental dynamics, and may offer new insights into the generic features of early-universe cosmology.

\section*{Acknowledgement}
SSH is supported in part by the National Institute for Theoretical
and Computational Sciences of South Africa (NITheCS).  The work of SD is supported by the Natural Sciences and Engineering Research Council of Canada. PKSD is supported by a grant from the First Rand Bank (SA). ST acknowledges support from the National Astrophysics and Space Science Programme (NASSP) at the University of Cape Town (UCT).

\bibliographystyle{unsrt}
\bibliography{main}

@article{rrnm-z3qy,
  author = {Das, S. and Dunsby, P. K. S. and Haque, S. S. and Tema, S.},
  title = {Power-law bounces in $f({R})$ gravity: analysis of the ekpyrosis and accelerating regimes},
  journal = {Phys. Rev. D},
  volume = {112},
  pages = {104059},
  year = {2025},
  doi = {10.1103/PhysRevD.112.104059}
}

@article{chakraborty2024model,
  author = {Chakraborty, S. and Louw, C. and Agrawal, A. S. and Dunsby, P. K. S.},
  title = {A model-independent compact dynamical system formulation for exploring bounce and cyclic cosmological evolutions in $f({R})$ gravity},
  journal = {Eur. Phys. J. C},
  volume = {84},
  pages = {1323},
  year = {2024},
  doi = {10.1140/epjc/s10052-024-1323}
}

@article{Guth:1980zm,
  author = {Guth, A. H.},
  title = {The inflationary universe: a possible solution to the horizon and flatness problems},
  journal = {Phys. Rev. D},
  volume = {23},
  pages = {347--356},
  year = {1981},
  doi = {10.1103/PhysRevD.23.347}
}

@article{Linde:1981mu,
  author = {Linde, A. D.},
  title = {A new inflationary universe scenario: a possible solution of the horizon, flatness, homogeneity, isotropy and primordial monopole problems},
  journal = {Phys. Lett. B},
  volume = {108},
  pages = {389--393},
  year = {1982},
  doi = {10.1016/0370-2693(82)91219-9}
}

@article{Dunsby:2015ers,
  author = {Dunsby, P. K. S. and Luongo, O.},
  title = {On the theory and applications of modern cosmography},
  journal = {Int. J. Geom. Meth. Mod. Phys.},
  volume = {13},
  pages = {1630002},
  year = {2016},
  doi = {10.1142/S0219887816300026},
  eprint = {1511.06532},
  archivePrefix = {arXiv},
  primaryClass = {gr-qc}
}

@article{Goliath:1998na,
  author = {Goliath, M. and Ellis, G. F. R.},
  title = {Homogeneous cosmologies with cosmological constant},
  journal = {Phys. Rev. D},
  volume = {60},
  pages = {023502},
  year = {1999},
  doi = {10.1103/PhysRevD.60.023502},
  eprint = {gr-qc/9811068},
  archivePrefix = {arXiv}
}

@article{Carloni:2004kp,
  author = {Carloni, S. and Dunsby, P. K. S. and Capozziello, S. and Troisi, A.},
  title = {Cosmological dynamics of ${R^n}$ gravity},
  journal = {Class. Quant. Grav.},
  volume = {22},
  pages = {4839--4868},
  year = {2005},
  doi = {10.1088/0264-9381/22/22/011},
  eprint = {gr-qc/0410046},
  archivePrefix = {arXiv}
}

@article{tHooft:1974kcl,
  author = {\'t Hooft, G.},
  title = {Magnetic monopoles in unified gauge theories},
  journal = {Nucl. Phys. B},
  volume = {79},
  pages = {276--284},
  year = {1974},
  doi = {10.1016/0550-3213(74)90486-6}
}

@article{Polyakov:1974ek,
  author = {Polyakov, A. M.},
  title = {Particle spectrum in quantum field theory},
  journal = {JETP Lett.},
  volume = {20},
  pages = {194--195},
  year = {1974}
}

@article{Preskill:1979zi,
  author = {Preskill, J.},
  title = {Cosmological production of superheavy magnetic monopoles},
  journal = {Phys. Rev. Lett.},
  volume = {43},
  pages = {1365},
  year = {1979},
  doi = {10.1103/PhysRevLett.43.1365}
}

@book{Weinberg:1972kfs,
  author = {Weinberg, S.},
  title = {Gravitation and cosmology: principles and applications of the general theory of relativity},
  publisher = {John Wiley and Sons},
  address = {New York},
  year = {1972}
}

@article{Busti:2015xqa,
  author = {Busti, V. C. and de la Cruz-Dombriz, A. and Dunsby, P. K. S. and S{\'a}ez-G{\'o}mez, D.},
  title = {Is cosmography a useful tool for testing cosmology?},
  journal = {Phys. Rev. D},
  volume = {92},
  pages = {123512},
  year = {2015},
  doi = {10.1103/PhysRevD.92.123512},
  eprint = {1505.05503},
  archivePrefix = {arXiv},
  primaryClass = {astro-ph.CO}
}

@article{Sandage:1961zz,
  author = {Sandage, A.},
  title = {The ability of the 200-inch telescope to discriminate between selected world models},
  journal = {Astrophys. J.},
  volume = {133},
  pages = {355--392},
  year = {1961},
  doi = {10.1086/147041}
}

@article{Albrecht:1982wi,
  author = {Albrecht, A. and Steinhardt, P. J.},
  title = {Cosmology for grand unified theories with radiatively induced symmetry breaking},
  journal = {Phys. Rev. Lett.},
  volume = {48},
  pages = {1220--1223},
  year = {1982},
  doi = {10.1103/PhysRevLett.48.1220}
}

@article{Starobinsky:1980te,
  author = {Starobinsky, A. A.},
  title = {A new type of isotropic cosmological models without singularity},
  journal = {Phys. Lett. B},
  volume = {91},
  pages = {99--102},
  year = {1980},
  doi = {10.1016/0370-2693(80)90670-X}
}

@article{PhysRevD.28.679,
  author = {Bardeen, J. M. and Steinhardt, P. J. and Turner, M. S.},
  title = {Spontaneous creation of almost scale-free density perturbations in an inflationary universe},
  journal = {Phys. Rev. D},
  volume = {28},
  pages = {679--693},
  year = {1983},
  doi = {10.1103/PhysRevD.28.679}
}

@article{Dicke:1965zz,
  author = {Dicke, R. H. and Peebles, P. J. E. and Roll, P. G. and Wilkinson, D. T.},
  title = {Cosmic black-body radiation},
  journal = {Astrophys. J.},
  volume = {142},
  pages = {414--419},
  year = {1965},
  doi = {10.1086/148306}
}

@book{Peebles:1994xt,
  author = {Peebles, P. J. E.},
  title = {Principles of physical cosmology},
  publisher = {Princeton University Press},
  year = {1994}
}

@article{Misner:1969hg,
  author = {Misner, C. W.},
  title = {Mixmaster universe},
  journal = {Phys. Rev. Lett.},
  volume = {22},
  pages = {1071--1074},
  year = {1969},
  doi = {10.1103/PhysRevLett.22.1071}
}

@article{Baumann:2018muz,
  author = {Baumann, D.},
  title = {Primordial cosmology},
  journal = {PoS TASI2017},
  pages = {009},
  year = {2018},
  doi = {10.22323/1.305.0009},
  eprint = {1807.03098},
  archivePrefix = {arXiv},
  primaryClass = {hep-th}
}

@article{Planck:2018vyg,
  author = {Aghanim, N. and others},
  title = {Planck 2018 results. VI. Cosmological parameters},
  journal = {Astron. Astrophys.},
  volume = {641},
  pages = {A6},
  year = {2020},
  doi = {10.1051/0004-6361/201833910},
  eprint = {1807.06209},
  archivePrefix = {arXiv},
  primaryClass = {astro-ph.CO}
}

@book{Kolb:1990vq,
  author = {Kolb, E. W. and Turner, M. S.},
  title = {The early universe},
  publisher = {Taylor and Francis},
  year = {1990},
  doi = {10.1201/9780429492860}
}

@article{Brandenberger:2016vhg,
  author = {Brandenberger, R. and Peter, P.},
  title = {Bouncing cosmologies: progress and problems},
  journal = {Found. Phys.},
  volume = {47},
  pages = {797--850},
  year = {2017},
  doi = {10.1007/s10701-016-0057-0},
  eprint = {1603.05834},
  archivePrefix = {arXiv}
}

@article{Lehners:2008vx,
  author = {Lehners, J.~L.},
  title = {Ekpyrotic and cyclic cosmology},
  journal = {Phys. Rept.},
  volume = {465},
  pages = {223--263},
  year = {2008},
  doi = {10.1016/j.physrep.2008.06.001},
  eprint = {0806.1245},
  archivePrefix = {arXiv}
}

@article{Kallosh:2013hoa,
  author = {Kallosh, R. and Linde, A.},
  title = {Universality class in conformal inflation},
  journal = {JCAP},
  volume = {07},
  pages = {002},
  year = {2013},
  doi = {10.1088/1475-7516/2013/07/002},
  eprint = {1306.5220},
  archivePrefix = {arXiv}
}

@article{Burger:2018hpz,
  author = {Burger, D. J. and Moynihan, N. and Das, S. and Haque, S. S. and Underwood, B.},
  title = {Towards the Raychaudhuri equation beyond general relativity},
  journal = {Phys. Rev. D},
  volume = {98},
  pages = {024006},
  year = {2018},
  doi = {10.1103/PhysRevD.98.024006},
  eprint = {1802.09499},
  archivePrefix = {arXiv}
}

@article{Das:2024rop,
  author = {Das, S. and Haque, S. S. and Tema, S.},
  title = {Cosmological singularity and power-law solutions in modified gravity},
  journal = {Annals Phys.},
  volume = {470},
  pages = {169829},
  year = {2024},
  doi = {10.1016/j.aop.2024.169829},
  eprint = {2405.09714},
  archivePrefix = {arXiv}
}

@article{Copeland:1997et,
  author = {Copeland, E. J. and Liddle, A. R. and Wands, D.},
  title = {Exponential potentials and cosmological scaling solutions},
  journal = {Phys. Rev. D},
  volume = {57},
  pages = {4686--4690},
  year = {1998},
  doi = {10.1103/PhysRevD.57.4686},
  eprint = {gr-qc/9711068},
  archivePrefix = {arXiv}
}

@article{Gasperini:1992em,
  author = {Gasperini, M. and Veneziano, G.},
  title = {Pre-big bang in string cosmology},
  journal = {Astropart. Phys.},
  volume = {1},
  pages = {317--339},
  year = {1993},
  doi = {10.1016/0927-6505(93)90017-8},
  eprint = {hep-th/9211021},
  archivePrefix = {arXiv}
}

@article{Copeland:2006wr,
  author = {Copeland, E. J. and Sami, M. and Tsujikawa, S.},
  title = {Dynamics of dark energy},
  journal = {Int. J. Mod. Phys. D},
  volume = {15},
  pages = {1753--1936},
  year = {2006},
  doi = {10.1142/S021827180600942X},
  eprint = {hep-th/0603057},
  archivePrefix = {arXiv}
}

@article{Peebles:1987ek,
  author = {Peebles, P. J. E. and Ratra, B.},
  title = {Cosmology with a time variable cosmological constant},
  journal = {Astrophys. J. Lett.},
  volume = {325},
  pages = {L17},
  year = {1988},
  doi = {10.1086/185100}
}

@article{Liddle:1998xm,
  author = {Liddle, A. R. and Scherrer, R. J.},
  title = {A classification of scalar field potentials with cosmological scaling solutions},
  journal = {Phys. Rev. D},
  volume = {59},
  pages = {023509},
  year = {1999},
  doi = {10.1103/PhysRevD.59.023509},
  eprint = {astro-ph/9809272},
  archivePrefix = {arXiv}
}

@article{Bahamonde:2017ize,
  author  = {S.~Bahamonde and C.~G.~B{\"o}hmer and S.~Carloni and E.~J.~Copeland and W.~Fang and N.~Tamanini},
  title   = {Dynamical systems applied to cosmology: dark energy and modified gravity},
  journal = {Phys.\ Rept.},
  volume  = {775-777},
  pages   = {1--122},
  year    = {2018},
  doi     = {10.1016/j.physrep.2018.09.001},
  eprint  = {1712.03107},
  archivePrefix = {arXiv},
  primaryClass  = {gr-qc}
}

@article{Vitagliano:2009sr,
  author  = {V.~Vitagliano and J.~Q.~Xia and S.~Liberati and M.~Viel},
  title   = {High-redshift cosmography},
  journal = {JCAP},
  volume  = {03},
  pages   = {005},
  year    = {2010},
  eprint  = {0911.1249}
}

@article{Damour:2002et,
  author  = {T.~Damour and M.~Henneaux and H.~Nicolai},
  title   = {Cosmological billiards},
  journal = {Class.\ Quant.\ Grav.},
  volume  = {20},
  pages   = {R145--R200},
  year    = {2003},
  eprint  = {hep-th/0212256}
}

@article{Spokoiny:1993kt,
  author  = {B.~Spokoiny},
  title   = {Deflationary universe scenario},
  journal = {Phys.\ Lett.\ B},
  volume  = {315},
  pages   = {40--45},
  year    = {1993}
}

@article{Eisenstein:2005su,
  author  = {D.~J.~Eisenstein et al.},
  title   = {Detection of the baryon acoustic peak in the large-scale correlation function of SDSS luminous red galaxies},
  journal = {Astrophys.\ J.},
  volume  = {633},
  pages   = {560--574},
  year    = {2005},
  eprint  = {astro-ph/0501171}
}

@article{Martin:2013tda,
  author  = {J.~Martin and C.~Ringeval and V.~Vennin},
  title   = {Encyclopaedia Inflationaris},
  journal = {Phys.\ Dark Univ.},
  volume  = {5-6},
  pages   = {75--235},
  year    = {2014},
  eprint  = {1303.3787}
}

@article{Roest:2013fha,
  author  = {D.~Roest},
  title   = {Universality classes of inflation},
  journal = {JCAP},
  volume  = {01},
  pages   = {007},
  year    = {2014},
  eprint  = {1309.1285}
}

@article{PhysRevLett.48.1437,
  title = {Reheating an Inflationary Universe},
  author = {A. Albrecht and P. J. Steinhardt and M. S. Turner and F. Wilczek},
  journal = {Phys. Rev. Lett.},
  volume = {48},
  pages = {1437--1440},
  year = {1982},
  doi = {10.1103/PhysRevLett.48.1437}
}

@article{Freese:1990rb,
  author    = {K. Freese and J. A. Frieman and A. V. Olinto},
  title     = {Natural inflation with pseudo Nambu-Goldstone bosons},
  journal   = {Phys. Rev. Lett.},
  volume    = {65},
  pages     = {3233--3236},
  year      = {1990},
  doi       = {10.1103/PhysRevLett.65.3233},
  url       = {https://link.aps.org/doi/10.1103/PhysRevLett.65.3233}
}

@article{Boubekeur:2005zm,
  author        = {L. Boubekeur and D. H. Lyth},
  title         = {Hilltop inflation},
  journal       = {JCAP},
  volume        = {07},
  pages         = {010},
  year          = {2005},
  doi           = {10.1088/1475-7516/2005/07/010},
  eprint        = {hep-ph/0502047},
  archivePrefix = {arXiv}
}

@book{Ryden:1970vsj,
    author = "B. Ryden",
    title = "{Introduction to Cosmology}",
    doi = "10.1017/9781316651087",
    isbn = "978-1-107-15483-4, 978-1-316-88984-8, 978-1-316-65108-7",
    publisher = "Cambridge University Press",
    year = "1970"
}

@book{Carroll:2004st,
    author = "S. M. Carroll",
    title = "{Spacetime and Geometry: An Introduction to General Relativity}",
    doi = "10.1017/9781108770385",
    isbn = "978-0-8053-8732-2, 978-1-108-48839-6, 978-1-108-77555-7",
    publisher = "Cambridge University Press",
    month = "7",
    year = "2019"
}

@article{PhysRevD.92.103004,
  author = "P.~H. Chavanis",
  title = "{Cosmology with a stiff matter era}",
  journal = "Phys. Rev. D",
  volume = "92",
  number = "10",
  pages = "103004",
  year = "2015",
  month = "11",
  doi = "10.1103/PhysRevD.92.103004",
  url = "https://link.aps.org/doi/10.1103/PhysRevD.92.103004"
}

@book{Peter:2013avv,
    author = "P. Peter and J.~P. Uzan",
    title = "{Primordial Cosmology}",
    isbn = "978-0-19-966515-0, 978-0-19-920991-0",
    publisher = "Oxford University Press",
    series = "Oxford Graduate Texts",
    month = "2",
    year = "2013"
}

@article{Senatore:2010wk,
    author = "L. Senatore and M. Zaldarriaga",
    title = "{The Effective Field Theory of Multifield Inflation}",
    eprint = "1009.2093",
    archivePrefix = "arXiv",
    primaryClass = "hep-th",
    doi = "10.1007/JHEP04(2012)024",
    journal = "JHEP",
    volume = "04",
    pages = "024",
    year = "2012"
}

@article{Sse,
    author = "D. A. Easson and B. A. Powell",
    title = "{The Degeneracy Problem in Non-Canonical Inflation}",
    eprint = "1212.4154",
    archivePrefix = "arXiv",
    primaryClass = "astro-ph.CO",
    doi = "10.1088/1475-7516/2013/03/028",
    journal = "JCAP",
    volume = "03",
    pages = "028",
    year = "2013"
}

@article{Aviles:2012ay,
    author = "A. Avilés and C. Gruber and O. Luongo and H. Quevedo",
    title = "{Cosmography and constraints on the equation of state of the Universe in various parametrizations}",
    eprint = "1204.2007",
    archivePrefix = "arXiv",
    primaryClass = "astro-ph.CO",
    doi = "10.1103/PhysRevD.86.123516",
    journal = "Phys. Rev. D",
    volume = "86",
    pages = "123516",
    year = "2012"
}

@article{Dunajski:2008tg,
    author = "M. Dunajski and G. W. Gibbons",
    title = "{Cosmic jerk, snap and beyond}",
    journal = "Class. Quant. Grav.",
    volume = "25",
    pages = "235012",
    year = "2008",
    doi = "10.1088/0264-9381/25/23/235012",
    eprint = "0807.0207",
    archivePrefix = "arXiv",
    primaryClass = "gr-qc",
    reportNumber = "DAMTP-2008-58"
}

@article{Brandenberger:2011eq,
  author = "R. H. Brandenberger",
  title = "{Is the spectrum of gravitational waves the ``Holy Grail'' of inflation?}",
  journal = "Eur. Phys. J. C",
  volume = "79",
  year = "2019",
  pages = "387",
  doi = "10.1140/epjc/s10052-019-6883-4"
}

\end{document}